\documentclass[prd,twocolumn,showpacs,amsmath,amssymb]{revtex4}
\usepackage{graphicx}
\begin{document}
\title{Dark viscous fluid described by a
unified equation of state in cosmology}
\author{Jie Ren$^1$}
\email{jrenphysics@hotmail.com}
\author{Xin-He Meng$^{2,3}$}
\email{xhm@nankai.edu.cn} \affiliation{$^1$Theoretical Physics
Division, Chern Institute of Mathematics, Nankai University, Tianjin
300071, China} \affiliation{$^2$Department of physics, Nankai
University, Tianjin 300071, China} \affiliation{$^3$Department of
physics, Hanyang University, Seoul 133-791, Korea}
\date{\today}
\begin{abstract}
We generalize the $\Lambda$CDM model by introducing a unified EOS to
describe the Universe contents modeled as dark viscous fluid,
motivated by the fact that a single constant equation of state (EOS)
$p=-p_0$ ($p_0>0$) reproduces the $\Lambda$CDM model exactly. This
EOS describes the perfect fluid term, the dissipative effect, and
the cosmological constant in a unique framework and the Friedmann
equations can be analytically solved. Especially, we find a relation
between the EOS parameter and the renormalizable condition of a
scalar field. We develop a completely numerical method to perform a
$\chi^2$ minimization to constrain the parameters in a cosmological
model directly from the Friedmann equations, and employ the SNe data
with the parameter $\mathcal{A}$ measured from the SDSS data to
constrain our model. The result indicates that the dissipative
effect is rather small in the late-time Universe.
\end{abstract}
 \maketitle

\textit{Introduction.} The cosmological observations have provided
increasingly convincing evidence that our Universe is undergoing a
late-time accelerating expansion \cite{rie98,bah99,ben03,spe06}, and
we live in a favored spatially flat Universe composed of
approximately $4\%$ baryonic matter, $22\%$ dark matter and $74\%$
dark energy. The simplest candidate for dark energy is the
cosmological constant. Recently, a great number of ideas have been
proposed to explain the current accelerating Universe, partly such
as scalar field model, exotic equation of state (EOS), modified
gravity, and the inhomogeneous cosmology model. However, the
available data sets in cosmology, especially the SNe Ia data
\cite{rie04,ast05,rie06}, the SDSS data \cite{eis05}, and the three
year WMAP data \cite{spe06} all indicate that the $\Lambda$CDM
model, which serves as a standard model in cosmology, is an
excellent model to describe the cosmological evolution. Therefore,
we suggest that a new cosmological model should be based on or can
be reduced to the $\Lambda$CDM model naturally.

Time-dependent bulk viscosity \cite{bre05}, a linear EOS
\cite{nai04,bab05}, and the Hubble parameter dependent EOS
\cite{noj05a} are considered in the study of the dark energy
physics. The EOS approach is intensely studied in cosmology, partly
such as Refs.~\cite{eos1,eos2,eos3,eos4,eos5,eos6,eos7,eos8,eos9}.
The equivalence between the modified EOS, the scalar field model,
and the modified gravity is demonstrated in
Refs.~\cite{cap05b,cap05c,noj05b}, with a general method to
calculate the potential of the corresponding scalar field for a
given EOS. We attempt to investigate the properties of cosmological
models starting from the EOS of the Universe contents directly,
which is suggested by the authors in Ref.~\cite{cap05a}. Our goal is
to find more physical meanings in the right hand side of the
Einstein equation to explore the currently accelerating universe. We
find that a generalized EOS unifies several issues in cosmology.
This EOS can be regarded as the generalization of the constant EOS
$p=-p_0$, which can reproduce the $\Lambda$CDM model exactly.

In this paper, we build up a general model called the extended
$\Lambda$CDM model, which can be exactly solved to describe the
cosmological evolutions by introducing a unified EOS. This paper is
a complement of our previous work \cite{ren06a,ren06b}, in which the
physical meanings of this EOS are very limited. We also develop a
completely numerical method to perform a $\chi^2$ minimization to
constrain the parameters of a cosmological model directly from the
Friedmann equations.

We consider the Friedmann-Robertson-Walker metric in the flat space
geometry ($k$=0) as the case favored by observational data
\begin{equation}
ds^2=-dt^2+a(t)^2(dr^2+r^2d\Omega^2),
\end{equation}
and assume that the cosmic fluid possesses a bulk viscosity $\zeta$.
The energy-momentum tensor is
\begin{equation}
T_{\mu\nu}=\rho U_\mu U_\nu+(p-\zeta\theta)h_{\mu\nu},
\end{equation}
where in comoving coordinates $U^\mu=(1,0,0,0)$,
$h_{\mu\nu}=g_{\mu\nu}+U_\mu U_\nu$, and $\theta=3\dot{a}/a$
\cite{bre02}. By defining the effective pressure as
$\tilde{p}=p-\zeta\theta$ and from the Einstein equation
$R_{\mu\nu}-\frac{1}{2}g_{\mu\nu}R=\kappa^2T_{\mu\nu}$, where
$\kappa^2=8\pi G$, we obtain the Friedmann equations
\begin{equation}
\frac{\dot{a}^2}{a^2} = \frac{\kappa^2}{3}\rho,~~\frac{\ddot{a}}{a}
= -\frac{\kappa^2}{6}(\rho+3\tilde{p}).\label{eq:eq1}
\end{equation}
The conservation equation for energy, $T^{0\nu}_{;\nu}=0$, yields
\begin{equation}
\dot{\rho}+3H(\rho+\tilde{p})=0,\label{eq:eq2}
\end{equation}
where $H=\dot{a}/a$ is the Hubble parameter.

\textit{Physical meaning of each term.} The EOS proposed in our
previous work \cite{ren06a} is given by
\begin{equation}
p=(\tilde{\gamma}-1)\rho-\frac{2}{\sqrt{3}\kappa
T_1}\sqrt{\rho}-\frac{2}{3\kappa^2 T_2^2}.\label{eq:eos}
\end{equation}
The first term is the prefect fluid EOS, the second term describes
the dissipative effect, and the third term corresponds to the
cosmological constant. The dynamical equation of the scale factor
$a(t)$ can be written as
\begin{equation}
\frac{\ddot{a}}{a}=-\frac{3\tilde{\gamma}-2}{2}\frac{\dot{a}^2}{a^2}
+\frac{1}{T_1}\frac{\dot{a}}{a}+\frac{1}{T_2^2}.\label{eq:main}
\end{equation}
The dimension of both $T_1$ and $T_2$ is [Time]. By concerning the
initial conditions of $a(t_0)=a_0$ and $\theta(t_0)=\theta_0$, the
analytical solution for $a(t)$ is given out in Ref.~\cite{ren06a}.

An intriguing feature of the extended $\Lambda$CDM model is that it
possesses both physical significance and mathematical exact
solutions. Form the physical point of view, Eq.~(\ref{eq:main})
naturally contains the dissipative process in the cosmological
evolution. If we set the EOS as $p=p_0$ and the bulk viscosity
coefficient $\zeta$ is constant, the first term describes the dark
matter, the last term ($T_2$ term) describes the dark energy, and
the middle term ($T_1$ term) describes the dissipative effects
probably coursed by the interaction between the dark matter and dark
energy. In Refs.~\cite{zim96,col96,chi97,pri00,bre04,cat05}, the
viscosity in cosmology has been studied in various aspects. The
qualitative analysis of Eq.~(\ref{eq:main}) can be easily obtained
if we assume that $H$ is always decreasing during the evolution of
the Universe. The three terms in the right-hand side of
Eq.~(\ref{eq:main}) are proportional to $H^2$, $H^1$, and $H^0$,
respectively, therefore, the three terms dominate alternatively
during the cosmological evolution and it approaches to a de Sitter
Universe finally. Actually, we can see that each term in the
right-hand side of Eq.~(\ref{eq:main}) accounts for the
time-dependent bulk viscosity or the variable cosmological constant.

\textit{Unified description of dark matter and dark energy.} The
$\Lambda$CDM model is based on the $H$-$z$ relation
\begin{equation}
H(z)^2=H_0^2[\Omega_m(1+z)^3+1-\Omega_m],\label{eq:lcdm}
\end{equation}
where $z=a_0/a-1$ is the redshift. We find that for a single
constant EOS $p=-p_0$ ($p_0>0$), the $H$-$z$ solution from the
Friedmann equations without viscosity is
\begin{equation}
H(z)^2=H_0^2\left[\left(1-\frac{\kappa^2p_0}{3H_0^2}\right)(1+z)^3
+\frac{\kappa^2p_0}{3H_0^2}\right],
\end{equation}
which exactly possesses the same form of Eq.~(\ref{eq:lcdm}), with
$\Omega_m=1-\frac{\kappa^2p_0}{3H_0^2}$. In the $\Lambda$CDM model,
the Universe contains two fluids, i.e., the dark matter and dark
energy, for which the EOS are $p=0$ and $p=-\rho$, respectively. In
our case, a single EOS unifies the dark matter and dark energy
modeled as dark viscous fluid, which is consistent with the
cosmological principle. However, it does not necessarily mean that
the nature of the dark matter and dark energy is the same. The
Chaplygin gas model $p=-A/\rho$ \cite{kam01} also serves as a
unified model of dark matter and dark energy, but it cannot reduce
to Eq.~(\ref{eq:lcdm}) exactly. As a special case of
Eq.~(\ref{eq:eos}), a linear EOS of the dark fluid is studied in
Ref.~\cite{nai04}, and the dark fluid is also studied by other
approaches, such as Ref.~\cite{arb05,arb06}.

Our motivation is to find a more general EOS, which possesses as
many as possible physical meanings and the Friedmann equations can
be exactly solved, as the following picture shows.
\begin{displaymath}
\framebox{$\left.\begin{array}{ll}
p=0 & (\textrm{CDM})\\
p=-\rho & (\Lambda)
\end{array} \right\} \Leftrightarrow p=-p_0$}
\rightarrow \framebox{Eq.~(\ref{eq:eos})}
\end{displaymath}
Luckily we obtain one which is just Eq.~(\ref{eq:eos}), and gives
rise to four implying unifications summarized at the end of this
article. Based on this EOS, we establish a cosmological model,
called the extended $\Lambda$CDM model.

\textit{Variable cosmological constant model.} It turns out that the
Friedmann equations combined with the renormalization equation which
determines the variable cosmological constant \cite{sha02a,sha02b}
can be reduced to the same form of Eq.~(\ref{eq:main})
\cite{ren06b}.

\textit{Scalar field model.} The authors of
Refs.~\cite{cap05b,noj05b} give a general method to obtain the
potential of a scalar model. We have found that the potential of the
corresponding scalar model is
\begin{equation}
V(\varphi)=C_1e^{\alpha\varphi}+C_2e^{\alpha\varphi/2}+C_3\label{eq:potential}
\end{equation}
if $\tilde{\gamma}\neq 0$ \cite{ren06b}. However, we missed an
important case, $\tilde{\gamma}=0$. In this case, the EOS is
\begin{equation}
p=-\rho-\frac{2}{\sqrt{3}\kappa T_1}\sqrt{\rho}-\frac{2}{3\kappa^2
T_2^2}.
\end{equation}
Using the same method, which is also outlined in Ref.~\cite{ren06b},
we obtain the potential of the corresponding scalar field
\begin{eqnarray}
V(\varphi) &=&
\frac{3\kappa^2}{64T_1^2}\varphi^4-\frac{3\kappa}{4\sqrt{2}T_1T_2}\varphi^3+
\left(\frac{3}{2T_2^2}-\frac{1}{8T_1^2}\right)\varphi^2\nonumber\\
&& +\frac{1}{\sqrt{2}\kappa T_1T_2}\varphi-\frac{1}{\kappa^2T_2^2}.
\end{eqnarray}
As a special case, if the bulk viscosity vanishes, $p=-\rho-p_0$.
The potential of the corresponding scalar field is
$V(\varphi)=\kappa^2p_0\varphi^2$ by neglecting the constant term.

In general, Eq.~(\ref{eq:potential}) is a non-renormalizable
potential, however, if the coefficient before $\rho$ is precisely
equal to $-1$, we obtain a renormalizable field. Moreover, the
$\sqrt{\rho}$ term in the EOS gives a contribution of $\varphi^4$
term in the scalar field. This property of such scalar field was
missed in our previous work. We think that there is a profound
relation between the renormalizability of the scalar field and that
the EOS parameter of the vacuum is precisely equal to $-1$.

\textit{Mathematical features.} In the mathematical aspect, the
transformation \cite{ren06b} $y=a^{3\tilde{\gamma}/2}$ reduces
Eq.~(\ref{eq:main}) to a linear differential equation of $y(t)$
\begin{equation}
\ddot{y}-\frac{1}{T_1}\dot{y}-\frac{3\tilde{\gamma}}{2T_2^2}y=0,
\end{equation}
which can be solved easily. The variable $y$ serves as a rescaled
scale factor and behaves like the amplitude of a damping harmonic
oscillator. The $T_1$ term is just the damping term. The equation
which determines the evolution of the Hubble parameter,
$\dot{H}=-\frac{3\tilde{\gamma}}{2}H^2+\frac{1}{T_1}H+\frac{1}{T_2^2}$,
has possessed a form invariance for $H\to H+\delta H$.

\textit{Supernovae constraints.} The observations of the SNe Ia have
provided the direct evidence for the cosmic accelerating expansion
for our current Universe. Any model attempting to explain the
acceleration mechanism should be consistent with the SNe Ia data
implying results, as a basic requirement. We have found the
viscosity without cosmological constant possesses a $(1+z)^{3/2}$
contribution \cite{ren06b}, which seems to be an interpolation
between the matter $(1+z)^3$ and the $\Lambda$-term $(1+z)^0$. The
method of the data fitting is illustrated in
Refs.~\cite{ben05,gon05}, in which the explicit solution $H(z)$ is
required. We develop a completely numerical method to perform a
$\chi^2$ minimization to fit the optimized values of the parameters
in a cosmological model directly from the Friedmann equations,
without knowing the $H$-$z$ relation. Define a new function
\begin{equation}
F(z)=\int_0^z\frac{dz}{E(z)},\label{eq:fz}
\end{equation}
where $E(z)=H(z)/H_0$ is the dimensionless Hubble parameter. The
relations implied by Eq.~(\ref{eq:fz})
\begin{equation}
E(z)=F'(z)^{-1},~~E'(z)=-F''(z)F'(z)^{-2}
\end{equation}
can transform an equation for $H(z)$ to another one for $F(z)$, then
one solves $F(z)$ numerically and obtains the luminosity distance
$d_L=(c/H_0)(1+z)F(z)$. This is a general numerical method and it
can be applied if only the dynamical equations determining the scale
factor is known.

\begin{figure}[]
\includegraphics{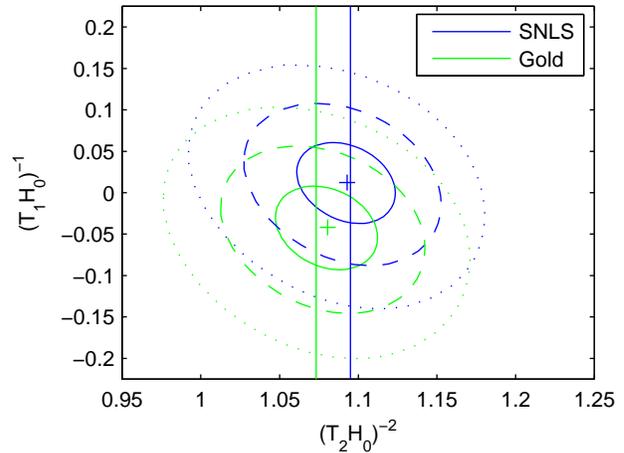}
\caption{\label{fig1} (color online). The $1\sigma$ (solid line),
$2\sigma$ (dashed line), and $3\sigma$ (dotted line) contour plots
of the $T_1$-$T_2$ relation in the extended $\Lambda$CDM model. The
vertical lines show the result of $T_2$ if it reduces to the
$\Lambda$CDM model ($T_1\to\infty$).}
\end{figure}

\begin{figure}[]
\includegraphics{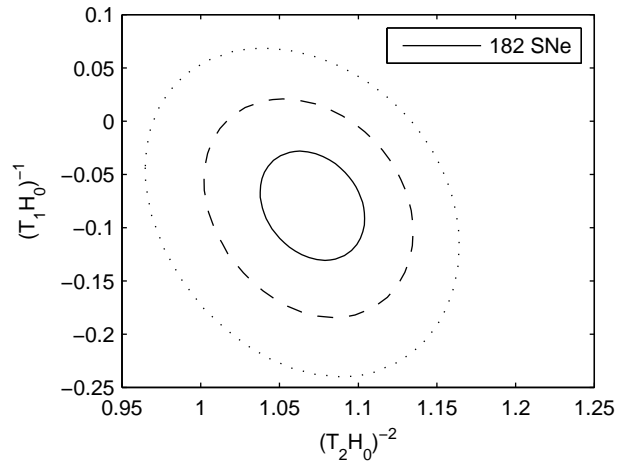}
\caption{\label{fig2} The $1\sigma$ (solid line), $2\sigma$ (dashed
line), and $3\sigma$ (dotted line) contour plots of the $T_1$-$T_2$
relation in the extended $\Lambda$CDM model.}
\end{figure}

The $\chi^2$ is calculated from
\begin{eqnarray}
\chi^2=\sum_{i=1}^{n}\left[\frac{\mu_{obs}(z_i)-\mathcal{M'}
-5\log_{10}D_{Lth}(z_i;c_\alpha)}{\sigma_{obs}(z_i)}\right]^2\nonumber\\
+\left(\frac{\mathcal{A}-0.469}{0.017}\right)^2,
\end{eqnarray}
where $\mathcal{M'}$ is a free parameter related to the Hubble
constant and $D_{Lth}(z_i;c_\alpha)$ is the theoretical prediction
for the dimensionless luminosity distance of a SNe Ia at a
particular distance, for a given model with parameters $c_\alpha$.
The parameter $\mathcal{A}$ is defined in Ref.~\cite{eis05}. Here
$\Omega_m=1-\frac{2}{3T_2^2H_0^2}$ is used in our model and we take
$\tilde{\gamma}=1$ as in the $\Lambda$CDM model. We will consider
the $\Lambda$CDM model for comparison and perform a best-fit
analysis with the minimization of the $\chi^2$, with respect to
$\mathcal{M'}$, $T_1H_0$, and $T_2H_0$. We employ the 157 gold data,
the SNLS data, and the 182 SNe data compiled by Riess {\it et al.}
recently combined with the parameter $\mathcal{A}$ to constrain the
parameters and plot the $T_1$-$T_2$ relation in Fig.~\ref{fig1} and
Fig.~\ref{fig2}. From the results, we see that the $T_1$ term is
made less than $10\%$ contributions to that of the $T_2$ term on
$2\sigma$ C.L. If we adopt the interpretation of viscosity of our
model, the fitting result shows that the dissipative effect is
rather small, as we expect that the additional term is a small
correction to the $\Lambda$CDM model.


\textit{Discussion.} The approach of the unified EOS considered in
this paper have enabled us to describe the Universe contents related
to several fundamental issues in cosmological evolution from a
united viewpoint. We have extended the $\Lambda$CDM model into a
more general framework by introducing this unified EOS. (i) This EOS
describes the perfect fluid term, the dissipative effect and the
cosmological constant in a unique equation. (ii) This general EOS
unifies the dark matter and the dark energy as a single dark viscous
fluid and can be exactly reduced to the $\Lambda$CDM model as a
special case. (iii) The variable cosmological constant model is
mathematically equivalent to the form by using this EOS. (iv) We
also find a scalar field that is equivalent to this EOS, moreover,
the renormalizable condition of the scalar field requires the
coefficient before $\rho$ is precisely equal to $-1$. Thus, it is
very interesting that concerning on the bulk viscosity, modified
EOS, variable cosmological constant model, and scalar field model
can be described in one general dynamical equation which determines
the scale factor. In this sense, our model has unified the exact
solutions of several models. The viewpoint of modified EOS is rather
phenomenological, however, we have showed that it is strongly
related to some fundamental concepts of cosmology. The incoming data
sets will give more constraints to the modified EOS approach in
cosmology.

X.-H.M. thank Prof. S.D. Odintsov for the helpful comments with
reading the manuscript, and Profs. I. Brevik and L. Ryder for lots
of discussions. X.-H.M is supported by the National Natural Science
Foundation of China (No. 10675062), and BK21 Foundation.


\end{document}